\documentclass[a4paper,fleqn]{cas-sc}

\usepackage[authoryear,longnamesfirst]{natbib}
\usepackage{subcaption}
\usepackage{siunitx}
\usepackage{amsmath,amssymb}
\usepackage{booktabs}
\usepackage{graphicx}
\usepackage{placeins}

\newcommand{\CD}{C_{\mathrm D}}

\newcommand{\E}{\mathbb{E}}
\newcommand{\Var}{\operatorname{Var}}
\newcommand{\Corr}{\operatorname{Corr}}
\newcommand{\Cov}{\operatorname{Cov}}

\newcommand{\ToyMomentBias}{1.0}
\newcommand{\ToyMomentNoise}{0.2}
\newcommand{\ToyResponseCorrelation}{0.981}
\newcommand{\ToySquaredCorrelation}{0.244}
\newcommand{\ToyLFRelativeCost}{0.1}
\newcommand{\ToyAllocationRatio}{15.8}
\newcommand{\ToyPerformanceRepeats}{5000}
\newcommand{\ToyLargestBudget}{320}
\newcommand{\ToyLargestMeanGain}{2.02}
\newcommand{\ToyLargestVariancePenalty}{1.59}
\newcommand{\ToyLargestMFMCNHigh}{123}
\newcommand{\ToyLargestMFMCNLow}{1944}

\begin{document}
\let\WriteBookmarks\relax
\def\floatpagepagefraction{1}
\def\textpagefraction{.001}

\shorttitle{Multifidelity Estimation of Drag-Coefficient Statistics}
\shortauthors{Boskovic et al.}

\title[mode=title]{Multifidelity Monte Carlo for Efficient Estimation of Drag-Coefficient Statistics in Rarefied Satellite Aerodynamics}

\author[1]{Jovan Boskovic}
\cormark[1]
\ead{jovan.boskovic@iag.uni-stuttgart.de}

\author[2]{Marcel Pfeiffer}
\ead{mpfeiffer@irs.uni-stuttgart.de}

\author[1]{Andrea Beck}
\ead{beck@iag.uni-stuttgart.de}

\affiliation[1]{organization={Institute for Aero- and Gasdynamics, University of Stuttgart},
            addressline={Wankelstraße 3},
            city={Stuttgart},
            postcode={70563},
            country={Germany}}

\affiliation[2]{organization={Institute of Space Systems, University of Stuttgart},
            addressline={Pfaffenwaldring 27},
            city={Stuttgart},
            postcode={70569},
            country={Germany}}

\cortext[1]{Corresponding author.}

\begin{abstract}
Very-low-Earth-orbit analyses require reliable uncertainty estimates for the aerodynamic drag coefficient, but repeated direct simulation Monte Carlo (DSMC) evaluations are expensive. We investigate whether cheaper collisionless models can reduce the computational cost of estimating DSMC-based drag statistics across different geometries. We also test whether improvements in the mean carry over to a variance reconstructed from two separately estimated raw moments. In the multifidelity Monte Carlo (MFMC) estimator, \textsc{PICLas} DSMC provides the high-fidelity response, while \textsc{PICLas} test-particle Monte Carlo (TPMC) and the free-molecular Sentman panel model provide low-fidelity control variates.

The benchmark contains four canonical Cube cases and scaled GOCE-like and CHAMP-like surface models. Each paired evaluation uses the same physical-input sample in all three models, and the reference values use 450 physically distinct DSMC evaluations per case. At the same estimation cost, DSMC--TPMC--Sentman MFMC reduces the RMSE of the mean drag coefficient in every tested case and budget, with median gains of 1.50--3.03. Variance estimation is less consistent: the three-model estimator improves three cases, performs close to HF-only Monte Carlo in two cases, and fails for GOCE-like. In that case, Sentman is only weakly correlated with the DSMC variance-related responses. Removing Sentman eliminates the negative estimates and improves the variance RMSE, although the decision to remove it was made after inspecting the results. MFMC therefore improves mean estimation reliably at reduced computational cost, but each low-fidelity model must be assessed separately for variance estimation.
\end{abstract}

\begin{highlights}
\item DSMC--TPMC--Sentman MFMC reduces mean-drag RMSE in all six benchmark cases.
\item Median mean-drag RMSE gains range from 1.50 to 3.03 at equal estimation cost.
\item Mean-response correlation does not ensure reliable variance reconstruction.
\item Removing Sentman post hoc stabilizes the GOCE-like variance estimate.
\end{highlights}


\maketitle

\section{Introduction}
\label{sec:intro}
Very low Earth orbit (VLEO) missions operate where aerodynamic drag affects orbit maintenance, lifetime prediction, and propellant budgeting \citep{walter2018astronautics,MostazaPrieto2014,Crisp2021}. In VLEO, the gas is rarefied, contains several species, and varies with solar and geomagnetic activity \citep{Picone2002,Emmert2015,Emmert2021}. Classical rarefied-gas and free-molecular theories provide the basis for relating this flow to satellite forces \citep{SchaafChambre1961,Sentman1961,Cook1965}. The aerodynamic response is represented here by the drag coefficient
\begin{equation}
  \CD(\boldsymbol{\xi},\omega)
  =\frac{F_{\mathrm D}(\boldsymbol{\xi},\omega)}
  {q_\infty(\boldsymbol{\xi})A_{\mathrm{ref}}},
  \qquad q_\infty=\frac{1}{2}\rho_\infty V_{\mathrm{rel}}^2,
  \label{eq:drag_coefficient_primary}
\end{equation}
Here $F_{\mathrm D}$ is drag force, $\rho_\infty$ is freestream mass density, $V_{\mathrm{rel}}$ is the gas--spacecraft relative speed, $q_\infty$ is dynamic pressure, $A_{\mathrm{ref}}$ is reference area, $\boldsymbol{\xi}$ contains uncertain physical inputs, and $\omega$ denotes numerical randomness. The resulting $\CD$ distribution is the primary quantity of interest.

Direct simulation Monte Carlo (DSMC) provides a kinetic high-fidelity (HF) model for rarefied flows \citep{Bird1994}. The open-source particle code \textsc{PICLas} supports DSMC \citep{Fasoulas2019} and has been applied to detailed DSMC simulations of satellite geometries in VLEO \citep{hild2022}. Conventional DSMC-only Monte Carlo is expensive because each additional physical-input sample requires a new stochastic particle calculation \citep{Mehta2023}. Collisionless TPMC and free-molecular panel methods such as the Sentman implementation in \textsc{ADBSat} are much cheaper \citep{Jin2019TPMC,Sentman1961,Sinpetru2022}. MFMC uses paired TPMC and Sentman evaluations as control variates, while DSMC continues to define the quantities being estimated.

The individual DSMC, TPMC, and free-molecular panel models are established \citep{Bird1994,Fasoulas2019,Jin2019TPMC,Sentman1961,Sinpetru2022}, but two questions remain when they are combined. First, can a DSMC--TPMC--Sentman hierarchy reduce the error in DSMC-based VLEO drag statistics at the same cost for both canonical and spacecraft-like geometries? Second, does a low-fidelity model that tracks $\CD$ also track $\CD^2$ and squared deviations from the mean? If not, an accurate mean estimate may still lead to a poor variance estimate when the variance is reconstructed from two separately corrected raw moments. The benchmark must therefore measure both computational efficiency and the correlations relevant to each statistic.

\paragraph{Purpose and contributions.}
We test whether DSMC--TPMC--Sentman and DSMC--TPMC estimators reduce the error in DSMC-based drag statistics at the same computational cost across six VLEO benchmark cases. We examine the mean and reconstructed variance separately because a model that tracks $\CD$ may not track $\CD^2$. We compare both multifidelity estimators with HF-only Monte Carlo, verify the implementation with analytic examples, and investigate why adding Sentman destabilizes the variance estimate in one case. The central result is that MFMC reliably improves the DSMC-based mean estimate, but the suitability of a low-fidelity model does not automatically carry over to nonlinear variance estimation.

Recent spacecraft applications show that MFMC can reduce the cost of uncertainty propagation in orbital dynamics and re-entry-window estimation \citep{SztamfaterGarcia2026}. That work combines dynamical propagators and estimates re-entry-time quantities. The present study considers an earlier step: molecular and free-molecular aerodynamic solvers are used to estimate the DSMC-based distribution of $\CD$. Its mean and uncertainty can then be passed to an orbit analysis instead of using one deterministic coefficient. Our contribution is limited to benchmarking DSMC, TPMC, and Sentman for drag statistics and diagnosing their performance for variance estimation. We do not claim a new general MFMC method or a new approach to spacecraft uncertainty propagation.

The study uses two low-fidelity branches. \textsc{PICLas} DSMC is the HF model, while \textsc{PICLas} TPMC and the \textsc{ADBSat} Sentman implementation are the LF models. All three receive the same sampled physical inputs. Section~\ref{sec:models} describes the gas--surface parameters and their model-specific implementation.

The benchmark is restricted to canonical Cube and scaled GOCE-like and CHAMP-like geometries. Detailed geometry models are important because errors in satellite shape and gas--surface assumptions propagate into derived drag and density products \citep{march2019geometry,march2020thesis}. These numerical cases test the estimators across different geometries and input distributions; they do not provide flight validation. We use only the surface models and reference-area conventions that passed the checks described below.

\paragraph{Paper organization.}
Section~\ref{sec:problem} defines the drag coefficient and input distributions. Section~\ref{sec:mfmc} presents the MFMC estimators for the two raw moments and explains how their correlations affect the reconstructed variance. Section~\ref{sec:models} describes the DSMC, TPMC, and Sentman models and their common gas--surface parameters. Section~\ref{sec:verification} verifies the implementation and defines the benchmark. Section~\ref{sec:results} reports the equal-cost results. Sections~\ref{sec:discussion}--\ref{sec:conclusion} interpret them, state the limitations, and answer the research question.

\section{Problem Formulation}
\label{sec:problem}
We first define the random inputs and the DSMC quantities whose mean and variance are estimated.
Let $k$ index a benchmark case and let $\boldsymbol{\xi}^{(k)}\in\Xi_k$ denote its complete random physical input with case-specific distribution $P_k$. For a sampled input, the high-fidelity response is
\begin{equation}
  f_0(\boldsymbol{\xi},\omega_0)
  =\widehat{C}_{\mathrm D}^{\mathrm{DSMC}}(\boldsymbol{\xi},\omega_0),
\end{equation}
where $\omega_0$ denotes the DSMC particle-sampling and other numerical random state. The two low-fidelity responses are
\begin{equation}
  f_1(\boldsymbol{\xi},\omega_1)=\widehat{C}_{\mathrm D}^{\mathrm{TPMC}}(\boldsymbol{\xi},\omega_1),
  \qquad
  f_2(\boldsymbol{\xi})=C_{\mathrm D}^{\mathrm{Sentman}}(\boldsymbol{\xi}).
\end{equation}
Passing the same $\boldsymbol{\xi}$ to all three models creates the paired responses used by the control variates. Both particle solvers have recorded numerical states, $\omega_0$ for DSMC and $\omega_1$ for TPMC; the Sentman evaluation is deterministic conditional on $\boldsymbol{\xi}$. Let $Q_k(\mathrm d\boldsymbol{\omega}\mid\boldsymbol{\xi})$, with $\boldsymbol{\omega}=(\omega_0,\omega_1)$, denote the conditional distribution of the numerical random variables induced by the fixed solver setups and seed procedures. The joint distribution is
\begin{equation}
  \Pi_k(\mathrm d\boldsymbol{\xi},\mathrm d\boldsymbol{\omega})
  =P_k(\mathrm d\boldsymbol{\xi})Q_k(\mathrm d\boldsymbol{\omega}\mid\boldsymbol{\xi}).
\end{equation}
The primary raw moments and reconstructed variance are
\begin{equation}
  \mu_0=\E_{\Pi_k}[f_0],\qquad
  m_{2,0}=\E_{\Pi_k}[f_0^2],\qquad
  \Var_{\Pi_k}[f_0]=m_{2,0}-\mu_0^2.
\end{equation}
With $m(\boldsymbol{\xi})=\E[f_0\mid\boldsymbol{\xi}]$ and $v_\omega(\boldsymbol{\xi})=\Var(f_0\mid\boldsymbol{\xi})$, the law of total variance gives
\begin{equation}
  \Var_{\Pi_k}[f_0]
  =\Var_{P_k}[m(\boldsymbol{\xi})]
  +\E_{P_k}[v_\omega(\boldsymbol{\xi})].
\end{equation}
The estimator uses one DSMC realization for each physical-input sample. Its variance therefore includes both physical-input variation and numerical variation. To measure the numerical variation at fixed inputs, we ran six independent DSMC seeds for nominal Cube and GOCE geometries at 300 km (Table~\ref{tab:dsmc-repeatability}). The corresponding standard deviations are $2.54\times10^{-4}$ and $1.63\times10^{-4}$, or 0.0276\% and 0.0196\% of the repeat means. Their sample variances are $1.45\times10^{-5}$ and $7.25\times10^{-7}$ of the corresponding benchmark variances. Particle noise is therefore small at these two fixed states. However, six repeats at one state cannot determine $\E_{P_k}[v_\omega(\boldsymbol{\xi})]$ over every case distribution. The variance studied here continues to include both physical and numerical variation.

\begin{table}[!htbp]
\centering
\caption{Fixed-input DSMC seed-repeatability diagnostic at 300 km. The joint standard deviation is taken from the corresponding 450-run benchmark distribution.}
\label{tab:dsmc-repeatability}
\begin{tabular}{lrrrr}
\toprule
Geometry & repeats & repeat SD($C_D$) & repeat SD [\%] & variance ratio \\
\midrule
Cube & 6 & 0.000254 & 0.0276 & 1.45e-05 \\
GOCE & 6 & 0.000163 & 0.0196 & 7.25e-07 \\
\bottomrule
\end{tabular}
\end{table}

\subsection{Benchmark Input Distributions}
\label{subsec:input_distribution}
The benchmark contains representative circular-orbit Cube cases and trajectory-conditioned scaled GOCE-like and CHAMP-like cases. Let $\boldsymbol{z}^{(k)}$ denote the explicitly configured continuous perturbations. For the Cube cases,
\begin{equation}
  \boldsymbol{z}^{(k)}=
  [\boldsymbol{z}_{\rho},\boldsymbol{z}_{x},\boldsymbol{z}_{T},
  \boldsymbol{z}_{w},\boldsymbol{z}_{a},\alpha_E,T_w]\in\mathbb{R}^{12}.
\end{equation}
The trajectory-conditioned cases omit separate configured wind perturbations and instead inherit the wind and environmental state through a sampled trajectory row $J^{(k)}$,
\begin{equation}
  \boldsymbol{z}^{(k)}_{\mathrm{traj}}=
  [\boldsymbol{z}_{\rho},\boldsymbol{z}_{x},\boldsymbol{z}_{T},
  \boldsymbol{z}_{a},\alpha_E,T_w]\in\mathbb{R}^{9},
  \qquad \boldsymbol{\xi}^{(k)}=(J^{(k)},\boldsymbol{z}^{(k)}_{\mathrm{traj}}).
\end{equation}
Here $\boldsymbol{z}_{\rho}$ contains density scale, temperature-offset, and composition-offset components; $\boldsymbol{z}_{x}$ and $\boldsymbol{z}_{T}$ are additional composition and freestream-temperature perturbations; $\boldsymbol{z}_{w}$ contains three wind components; and $\boldsymbol{z}_{a}=(\beta,\alpha)$ contains sideslip and angle of attack. The trajectory row jointly supplies time, position, altitude, solar and geomagnetic indices, density, composition, temperature, relative speed, and the state required for the HWM14 wind evaluation.

\begin{table}[!htbp]
\centering
\caption{Configured continuous input distributions for the Cube and trajectory-conditioned cases. Normal draws are clipped to the stated intervals. The trajectory environment is additionally conditioned on a discrete empirical row draw.}
\label{tab:production_sources}
\small
\begin{tabular}{lll}
\toprule
Input & Cube & GOCE-like/CHAMP-like trajectory cases \\
\midrule
Density scale & $\mathcal{N}(1,0.1^2)$; $[0.7,1.5]$ & same \\
Density-state temperature offset [K] & $\mathcal{N}(0,60^2)$; $[-200,200]$ & same \\
Density-state composition shift & $\mathcal{N}(0,0.04^2)$; $[-0.18,0.18]$ & same \\
Composition shift & $\mathcal{N}(0,0.05^2)$; $[-0.25,0.25]$ & same \\
Temperature offset [K] & $\mathcal{N}(0,100^2)$; $[-300,300]$ & same \\
Wind components & configured normal perturbations & trajectory row/HWM14 \\
AoS and AoA [deg] & $\mathcal{N}(0,2^2)$; $[-5,5]$ & same \\
Energy accommodation $\alpha_E$ & $\mathcal{N}(0.9,0.05^2)$; $[0.75,1]$ & same \\
Wall temperature $T_w$ [K] & $\mathcal{N}(300,25^2)$; $[250,380]$ & same \\
\bottomrule
\end{tabular}
\end{table}

These distributions are generic engineering choices used to test the estimator; they were not fitted to data. The density, composition, and temperature perturbations are centered on the configured NRLMSIS states. Their broad clipped ranges represent known thermospheric variability and differences between atmospheric density models \citep{Emmert2015,Emmert2021,sagnieres2017}, and HWM14 provides the wind state \citep{Drob2015HWM14}. The incidence-angle and wall-state ranges define bounded operating conditions. The energy-accommodation range represents a simplified family of VLEO surfaces with high accommodation, not a posterior distribution for a specific material. Previous VLEO studies show that aerodynamic coefficients depend on the atmospheric state, incidence, and gas--surface accommodation. In-orbit accommodation depends on the surface state and environment and is not known precisely for an arbitrary surface \citep{Moe2005}. We sample the listed quantities independently to obtain a reproducible test distribution. This distribution is not an empirically fitted joint model of the environment.

\paragraph{Gas--surface interaction uncertainty.}
The wall interaction separates fixed diffuseness from uncertain energy exchange. The \textsc{PICLas} momentum-accommodation coefficient is fixed at
\begin{equation}
  \alpha_{\mathrm{mom}}^{\mathrm{PICLas}}=1,
\end{equation}
which defines fully diffuse momentum reflection for every sample. Vibrational and rotational accommodation are also fixed at unity. The PICLas particle-boundary documentation defines separate coefficients for momentum reflection and translational, rotational, and vibrational energy accommodation \citep{PICLasDocs2026}. The scalar $\alpha_E$ remains an uncertain physical input and is mapped samplewise as
\begin{equation}
  \mathrm{TransACC}^{\mathrm{PICLas}}=\alpha_E,
  \qquad
  \alpha_{\mathrm{Sentman}}=\alpha_E.
  \label{eq:gsi_mapping}
\end{equation}
Thus, ``fully diffuse'' does not mean that $\alpha_E$ is fixed to one. It describes the angular and momentum reflection, whereas $\alpha_E$ controls translational energy accommodation. DSMC, TPMC, and Sentman all receive the same sampled values of $\alpha_E$ and $T_w$; no additional surface uncertainty is applied only to the low-fidelity models. The distribution of $\alpha_E$ is a generic modelling choice and was not fitted to data for a specific material.

Within each case, the configured continuous quantities are sampled independently and then clipped. Normal draws outside a stated bound are set to that bound rather than redrawn. The resulting distributions therefore contain probability mass at the bounds. Derived quantities for atmosphere, composition, speed, and incidence can still be dependent because they share the same baseline state. Variables taken from a trajectory row also keep their empirical dependence. Density affects $\CD$ indirectly by changing collisionality, composition, temperature, and speed ratio. Dynamic pressure multiplies $\CD$ only when calculating the derived quantity $p_{\mathrm D}$.

\paragraph{Monte Carlo propagation.}
A draw from $P_k$ defines the complete atmosphere, relative flow, incidence, $\alpha_E$, and $T_w$. All fidelities receive that same physical draw. DSMC additionally receives a sample-specific numerical seed. The resulting uncertainty metrics are case-specific moments of $\CD$ under the corresponding joint distribution; they are not moments under one universal cross-case distribution.

\section{Multifidelity Estimation of Mean and Variance}
\label{sec:mfmc}
MFMC reduces the number of DSMC evaluations by using correlated TPMC and Sentman predictions to correct the DSMC sample mean. Each low-fidelity model is evaluated on the DSMC inputs and on additional inputs. The difference between its means over these two sample sets has an expected value of zero and serves as a control-variate correction. The sets are nested: every input evaluated by an expensive model is also evaluated by each cheaper model. This provides paired outputs on the shared inputs and allows many additional inexpensive evaluations. We combine these corrections with DSMC using the MFMC method of \citet{Peherstorfer2016}. Because the drag-coefficient variance is calculated from two raw moments, we estimate the mean and second raw moment with separate weights.

For the mean estimator, consider nested sample sets
\begin{equation}
  \mathcal{S}_0\subset\mathcal{S}_1\subset\cdots\subset\mathcal{S}_{L_f},\qquad
  |\mathcal{S}_\ell|=N_\ell,\qquad N_0\le N_1\le\cdots\le N_{L_f}.
\end{equation}
Here $\mathcal{S}_\ell$ is the set of input realizations evaluated by model $\ell$, and $N_\ell$ is its sample count. Nesting ensures that every HF sample is paired with all LF outputs used in its correction.
To keep the estimator equations short, $f_\ell(\boldsymbol{\xi}_j)$ denotes $f_\ell(\boldsymbol{\xi}_j,\omega_{\ell j})$ for the particle models $\ell=0,1$ and the deterministic Sentman value. For any model $f_\ell$, define the sample mean over the first $N$ paired samples by
\begin{equation}
  \overline{f}_\ell^{(N)}=\frac{1}{N}\sum_{j=1}^{N}f_\ell(\boldsymbol{\xi}_j).
\end{equation}
The MFMC estimator for the high-fidelity mean is
\begin{equation}
  \label{eq:mfmc_mean}
  \widehat{\mu}_0=\overline{f}_0^{(N_0)}
  +\sum_{\ell=1}^{L_f}w_\ell\left(\overline{f}_\ell^{(N_\ell)}
  -\overline{f}_\ell^{(N_{\ell-1})}\right).
\end{equation}
Under independent and identically distributed sampling, each low-fidelity correction has an expected value of zero. Therefore, $\widehat{\mu}_0$ is unbiased for $\mu_0$ when the weights and sample counts are fixed. The standard nested MFMC estimator uses one scalar control-variate weight for each low-fidelity model \citep{Peherstorfer2016}. The optimal weight for the population is
\begin{equation}
  w_\ell^{\star}
  =\frac{\Cov(f_0,f_\ell)}{\Var(f_\ell)}
  =\rho_{f,\ell}\frac{\sigma_0}{\sigma_\ell},
  \qquad \ell=1,\ldots,L_f,
\end{equation}
where $\sigma_0$ and $\sigma_\ell$ are the standard deviations of the HF response and the corresponding LF response. In this study, $L_f=2$. We estimate the weights from pilot samples evaluated with DSMC, TPMC, and Sentman. The pilot and estimator samples are disjoint, so the fitted weights are independent of the estimator samples. Before estimation, we resample the pilot to check the stability of the weights and sample counts.
The sample counts are chosen under the cost constraint
\begin{equation}
  \sum_{\ell=0}^{L_f}c_\ell N_\ell\le B,\qquad N_0\le N_1\le\cdots\le N_{L_f},
\end{equation}
where $c_\ell$ is the measured cost of one evaluation of model $\ell$ and $B$ is the total computational budget. Costs and budgets are reported in DSMC-equivalent evaluations, obtained by dividing by the cost of one DSMC evaluation. Pilot estimates of costs and covariances determine the feasible allocation.
For the general nested estimator, order the LF models by the absolute pilot correlations between their drag coefficients and the DSMC drag coefficient, $|\rho_{C,1}|\ge\cdots\ge|\rho_{C,L_f}|$, and set $\rho_{C,L_f+1}=0$. The continuous standard-MFMC count ratios are
\begin{equation}
  r_\ell
  =\sqrt{\frac{c_0\left(\rho_{C,\ell}^2-\rho_{C,\ell+1}^2\right)}
  {c_\ell\left(1-\rho_{C,1}^2\right)}},
  \qquad N_\ell\approx r_\ell N_0.
\end{equation}
Here $\rho_{C,\ell}=\Corr(\CD^{\mathrm{DSMC}},\CD^{(\ell)})$ is the pilot correlation between the DSMC drag coefficient and the corresponding low-fidelity drag coefficient. For numerical stability, the implementation limits squared correlations to values below one. It also requires nondecreasing count ratios and sample counts, applies the configured maximum, and chooses the largest affordable integer $N_0$ for the measured-cost budget. The resulting sample counts are therefore a feasible, rounded approximation to the continuous formula. We use these counts for both raw moments but fit separate scalar weights for each moment.

\subsection{Second Raw Moment}
\label{subsec:second_raw_moment}
The second raw moment is estimated as a separate high-fidelity target. We define the squared responses
\begin{equation}
  g_\ell(\boldsymbol{\xi})=f_\ell(\boldsymbol{\xi})^2,\qquad \ell=0,\ldots,L_f.
\end{equation}
The second-moment estimator uses squared responses and separate control-variate weights $\widetilde{w}_\ell$. It uses the same nested sample sets and sample counts as the mean estimator, $\mathcal{T}_\ell=\mathcal{S}_\ell$ and $M_\ell=N_\ell$:
\begin{equation}
  \widehat{m}_{2,0}=\overline{g}_0^{(M_0)}
  +\sum_{\ell=1}^{L_f}\widetilde{w}_\ell\left(\overline{g}_\ell^{(M_\ell)}
  -\overline{g}_\ell^{(M_{\ell-1})}\right).
\end{equation}
We estimate $\widetilde{w}_\ell$ from the pilot covariances between squared drag-coefficient responses; we do not reuse the weights from the mean estimator. By contrast, \citet{qian2018multifidelity} construct a direct, unbiased MFMC variance estimator from nested sample variances,
\begin{equation}
  \widehat{V}_{\mathrm{MF}}=\widehat{V}_{0}^{(N_0)}
  +\sum_{\ell=1}^{L_f}\alpha_\ell
  \left(\widehat{V}_{\ell}^{(N_\ell)}-\widehat{V}_{\ell}^{(N_{\ell-1})}\right),
\end{equation}
where $\widehat{V}_{\ell}^{(N)}$ is the unbiased sample variance of model $\ell$ over $N$ nested inputs. They also show that optimizing this variance estimator directly requires uncertain fourth-order statistics from the pilot, and they recommend choosing sample counts based on the mean. We likewise choose the common sample counts for both raw moments from the drag-coefficient mean. We then fit separate control-variate weights for the mean and second raw moment. We keep these sample counts fixed and do not claim that they are optimal for the mean, the second raw moment, or the reconstructed variance. We reconstruct the drag-coefficient variance from the two raw-moment estimators instead of applying the direct Qian variance estimator. The completed simulation campaign used common sample counts selected from the mean response and stored the paired outputs required for both raw moments. We retain this design so that all estimators use the same simulations and can be compared at the same cost.

\subsection{Reconstructing and Assessing the Variance}
The reconstructed variance is more sensitive to error than either raw-moment estimate alone. It is the difference between two quantities of similar size. Small errors in $\widehat{m}_2$ and $\widehat{\mu}^{2}$ can therefore dominate the result, especially when they act in opposite directions. A low-fidelity model can reduce the error in the mean yet worsen the final variance if its correlation changes after squaring or if the errors in the two corrected raw moments do not cancel as required.

After estimating the two raw moments, the variance is reconstructed as
\begin{equation}
  \widehat{\Var}_{\Pi_k}[\CD]=\widehat{m}_{2,0}-\widehat{\mu}_0^2.
\end{equation}
This is a nonlinear estimator derived from the two raw-moment estimates. Even if both raw-moment estimators are unbiased for fixed weights, $\widehat{\mu}_0^2$ generally introduces finite-sample bias. We therefore assess the variance estimator by its RMSE relative to a DSMC-only reference and do not claim that it is unbiased.

For each low-fidelity control variate, the relevant diagnostic correlations are
\begin{equation}
  \rho_{f,\ell}=\operatorname{corr}(f_0,f_\ell),\qquad
  \rho_{g,\ell}=\operatorname{corr}(f_0^2,f_\ell^2).
\end{equation}
The first correlation indicates whether the low-fidelity model can reduce error in $\E[\CD]$; the second does the same for $\E[\CD^2]$. Both are needed to assess the reconstructed variance. A large $\rho_{f,\ell}$ and a much smaller $\rho_{g,\ell}$ warn that the mean estimate may improve while the second-moment and variance estimates remain unreliable. Section~\ref{sec:verification} gives an analytic example in which $\rho_f$ is close to one but $\rho_g$ is small.

Because variance is a centered second moment, we additionally inspect
\begin{equation}
  \rho_{v,\ell}=\operatorname{corr}\!\left(
  (f_0-\overline f_0)^2,(f_\ell-\overline f_\ell)^2\right)
\end{equation}
on the paired pilot samples. We use this correlation only to assess the model, not to choose the sample counts. A low $\rho_{v,\ell}$ shows that separate corrections to the two raw moments may not preserve the cancellation in $\widehat m_2-\widehat\mu^2$. If we remove a model after inspecting the estimation results, we report that analysis as exploratory rather than as a prespecified comparison.

For a single LF branch considered in isolation with its optimal scalar control-variate weight, the ideal residual variance fraction is \citep{Peherstorfer2016}
\begin{equation}
  \frac{\Var[f_0-w_\ell^\star(f_\ell-\E[f_\ell])]}{\Var[f_0]}
  =1-\rho_{f,\ell}^2.
\end{equation}
We therefore report the correlations for both the response and its square. We use $1-\rho_{f,1}^2$ only to show the ideal fraction of variance that would remain with one optimally weighted control variate. It does not predict performance at a finite budget, which also depends on the sample counts, error in the pilot estimates, and measured model costs. We calculate this quantity separately for $\CD$ and $\CD^2$.

A first-order delta-method expansion \citep{Oehlert1992} makes the cancellation sensitivity explicit. Let $\Delta_\mu=\widehat{\mu}-\mu$ and $\Delta_m=\widehat{m}_2-m_2$. To first order,
\begin{equation}
  \widehat{\Var}_{\Pi_k}[\CD]-\Var_{\Pi_k}[\CD]
  \approx \Delta_m-2\mu\Delta_\mu,
\end{equation}
so that
\begin{equation}
  \operatorname{Var}\!\left(\widehat{\Var}_{\Pi_k}[\CD]\right)
  \approx \operatorname{Var}(\widehat{m}_2)
  +4\mu^2\operatorname{Var}(\widehat{\mu})
  -4\mu\,\operatorname{Cov}(\widehat{m}_2,\widehat{\mu}).
\end{equation}
Variance gains therefore depend not only on the two raw-moment errors, but also on their covariance.

\subsection{When MFMC May Fail}
Using separate MFMC weights for each moment can fail or provide little benefit in several situations: the LF/HF correlation is low for $\CD$ or $\CD^2$; a source of uncertainty changes how the models respond relative to each other; the pilot covariances are unstable; the LF models are not cheap enough relative to DSMC; or the target variance is small compared with the two raw moments. For reference, one optimally weighted control variate reduces the variance of the corrected mean to a fraction $1-\rho^2$ of the uncorrected HF variance.

This scaling shows why the pilot correlations must be checked before using a low-fidelity model. Even a moderate decrease in correlation can remove most of the MFMC gain. A cheap and deterministic model can still be a poor control variate if it responds to changes in atmosphere, incidence direction, or surface state differently from DSMC. Conversely, a biased low-fidelity model can be useful if its changes across the sampled inputs follow those of DSMC.

\section{Physical and Numerical Models}
\label{sec:models}
The simulation hierarchy contains three models: \textsc{PICLas} DSMC as the high-fidelity target, collisionless \textsc{PICLas} TPMC as the intermediate-fidelity model, and the \textsc{ADBSat} Sentman panel model as the cheapest control variate.

\paragraph{Direct Simulation Monte Carlo in \textsc{PICLas}.}
\textsc{PICLas} DSMC provides the stochastic high-fidelity drag response \citep{Bird1994,Fasoulas2019}. The unstructured meshes used for these simulations were generated with \textsc{PyHOPE} \citep{Kopper2025PyHOPE}. The force convention is
\begin{equation}
  \vec{F}=\sum_j\vec{t}_jA_j,
  \qquad
  \CD=\frac{-\vec{F}\cdot\widehat{\mathbf e}_{V}}
  {q_\infty A_{\mathrm{ref}}},
  \label{eq:drag_coefficient}
\end{equation}
where $\widehat{\mathbf e}_{V}$ is the freestream direction and $A_{\mathrm{ref}}$ is a fixed reference area for each geometry. We include new solver outputs only after the corresponding Cube, GOCE-like, or CHAMP-like mesh and reference area pass checks of geometry, surface-normal orientation, projected area, and force sign.

\paragraph{PICLas wall model.}
The wall boundary uses fully diffuse momentum reflection, with `MomentumACC` fixed to one. Translational accommodation remains uncertain: the sampled $\alpha_E$ is assigned to `TransACC` according to Eq.~\eqref{eq:gsi_mapping}. Vibrational and rotational accommodation are fixed at unity, and every paired evaluation uses the sampled wall temperature. Keeping these parameters separate ensures that the diffuse-reflection setting does not remove the intended uncertainty in gas--surface energy exchange.

\paragraph{Test-particle Monte Carlo.}
The intermediate-fidelity model uses the same \textsc{PICLas} geometry, reference area, sampled atmosphere, incidence, and wall parameters as DSMC, but omits intermolecular collisions. It retains particle-based surface interactions and geometric visibility at a lower cost than DSMC, consistent with the standard TPMC formulation for satellite aerodynamics \citep{Jin2019TPMC}. TPMC serves only as a control variate; DSMC still defines the quantity being estimated.

\paragraph{Sentman low-fidelity model.}
\textsc{ADBSat} evaluates the classical Sentman equations for diffuse re-emission in free-molecular flow on the corrected triangulated surface \citep{Sentman1961,Sinpetru2022}. The contribution of each surface element depends on incidence, speed ratio, wall temperature, and the sampled $\alpha_E$. The total drag coefficient uses the same reference area and freestream convention as Eq.~\eqref{eq:drag_coefficient}. Shadowing and exposure filters are enabled for complex geometries, following the ADBSat shading treatment \citep{Sinpetru2022}. Sentman is used only as a control variate. A correlation with DSMC does not by itself show that its absolute aerodynamic prediction is valid.

\paragraph{Numerical target and quality checks.}
The DSMC response remains stochastic even when $\boldsymbol{\xi}$ is fixed. We assess numerical accuracy separately from variation in the physical inputs by repeating fixed inputs with different seeds, varying the particle number and time step, and checking the sampling window, mesh, and domain. We use only outputs from the corrected GOCE-like and CHAMP-like surfaces with the checked reference areas. Results from the earlier surface models are excluded rather than corrected after the simulations.

\section{Verification and Benchmarking}
\label{sec:verification}
We test both the numerical implementation and its computational efficiency. First, an analytic example with closed-form moments verifies the MFMC implementation. We then compare MFMC with DSMC-only Monte Carlo for the mean and variance of the drag coefficient under the input distributions defined in Sec.~\ref{subsec:input_distribution}.

\subsection{Numerical Verification: Analytic Toy Model}
\begin{figure}[!htbp]
  \centering
  \includegraphics[width=0.8\textwidth]{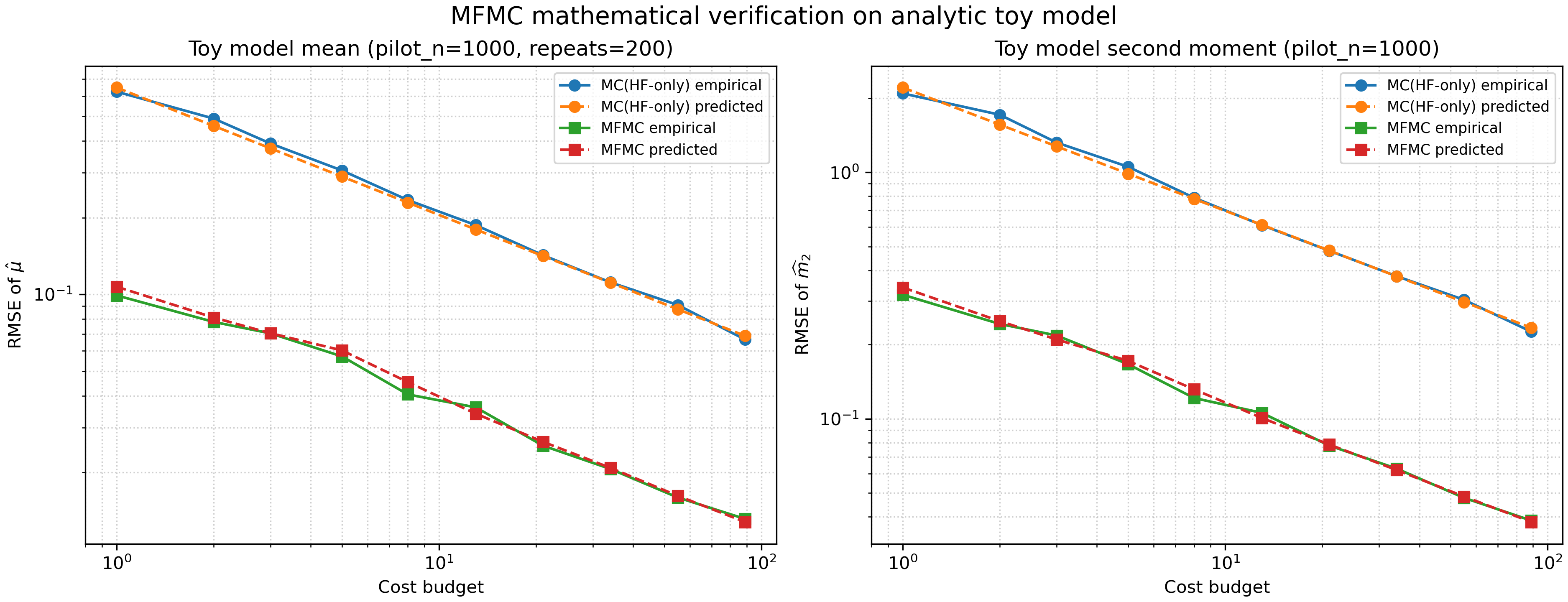}
  \caption{Analytic toy-model verification: empirical versus predicted RMSE of $\widehat{\mu}$ and $\widehat{m}_2$ versus total cost for HF-only Monte Carlo and MFMC.}
  \label{fig:toy_model}
\end{figure}

We define $X\sim\mathcal{U}(-1,1)$ and a high-fidelity quantity
\begin{equation}
  Y_H=f(X),\qquad f(X)=1+X+X^2,
\end{equation}
with two low-fidelity control-variate models
\begin{equation}
  Y_{L1}=f(X)+aX^3,\qquad Y_{L2}=f(X)+bX^5,\qquad (a,b)=(0.2,0.4).
\end{equation}
For this construction, $\E[Y_H]=4/3$ and $\E[Y_H^2]=11/5$. Synthetic per-model costs are fixed to $w_H=1$, $w_{L1}=0.05$, and $w_{L2}=0.01$. Pilot statistics use $N_{\mathrm{pilot}}=1000$ shared samples, and empirical RMSE curves use $R=200$ independent repeats. Figure~\ref{fig:toy_model} compares the predicted and empirical errors over the tested budgets; their close agreement verifies the implementation for both raw-moment targets.

\begin{figure}[!htbp]
  \centering
  \includegraphics[width=0.82\textwidth]{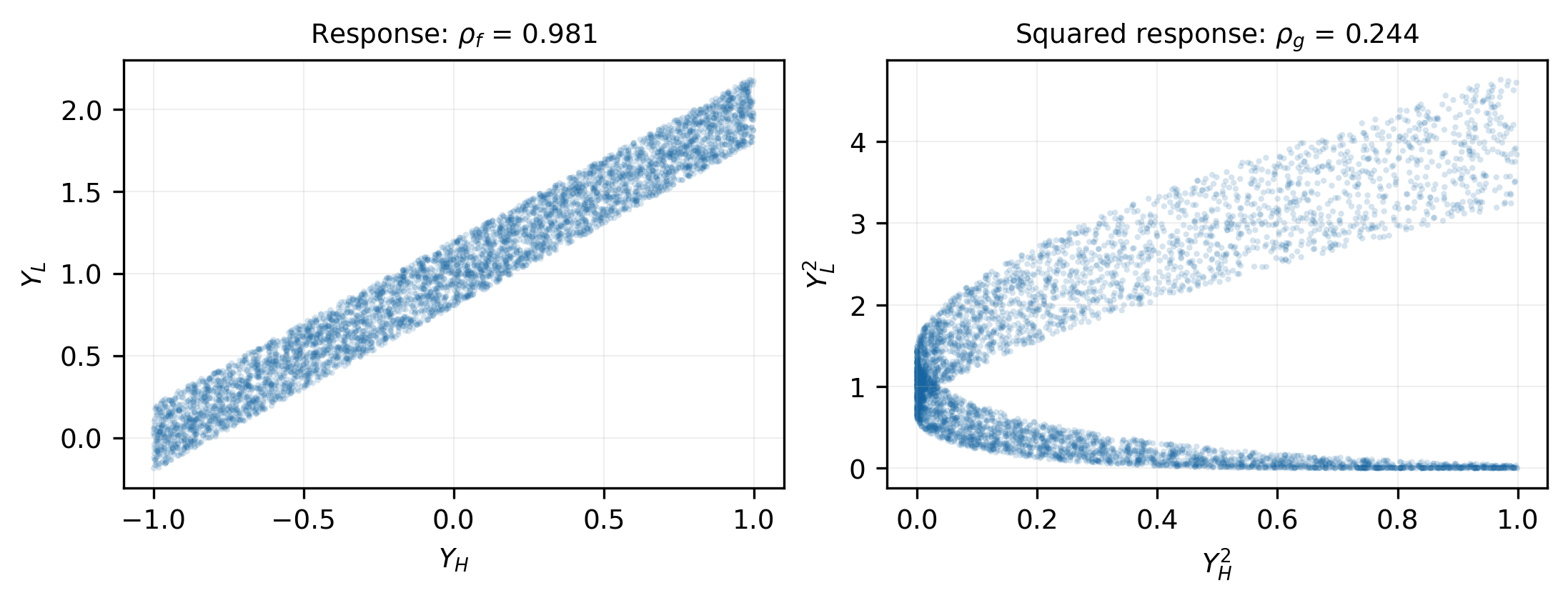}
  \caption{Analytic moment-correlation counterexample. The same paired samples form a narrow, strongly correlated band for the unsquared responses (left), whereas the squared responses are only weakly correlated (right). Correlation values in the panel titles are the analytic population values; points are a seeded visualization.}
  \label{fig:toy_moment_correlation}
\end{figure}

\paragraph{Moment-correlation counterexample.}
The following example isolates the main risk studied in this paper: a low-fidelity model can be strongly correlated with the HF response but weakly correlated with its square. In the preceding example, the response and its square have similarly strong correlations between the HF and LF models. A second analytic example shows that this need not be the case. Let $X,Z\sim\mathcal{U}(-1,1)$ be independent and define
\begin{equation}
  Y_H=X,\qquad Y_L=X+b+\varepsilon Z,\qquad (b,\varepsilon)=(\ToyMomentBias,\ToyMomentNoise).
\end{equation}
The additive LF bias $b$ disappears when the unsquared responses are centered. Consequently,
\begin{equation}
  \operatorname{corr}(Y_H,Y_L)=\frac{1}{\sqrt{1+\varepsilon^2}}=\ToyResponseCorrelation.
\end{equation}
After squaring, however, the term $2b(X+\varepsilon Z)$ contributes variance to $Y_L^2$ without corresponding covariance with $Y_H^2$. Evaluation of the exact uniform moments gives
\begin{equation}
  \operatorname{corr}(Y_H^2,Y_L^2)=\ToySquaredCorrelation.
\end{equation}
Figure~\ref{fig:toy_moment_correlation} shows the difference between the two correlations. An LF model can therefore be an excellent linear control variate for the first raw moment but a weak control variate for the second raw moment. This abstract example shows only that $\rho_f\approx1$ does not imply $\rho_g\approx1$; it does not claim that the same difference occurs in the drag pilots.

\begin{figure}[!htbp]
  \centering
  \includegraphics[width=0.82\textwidth]{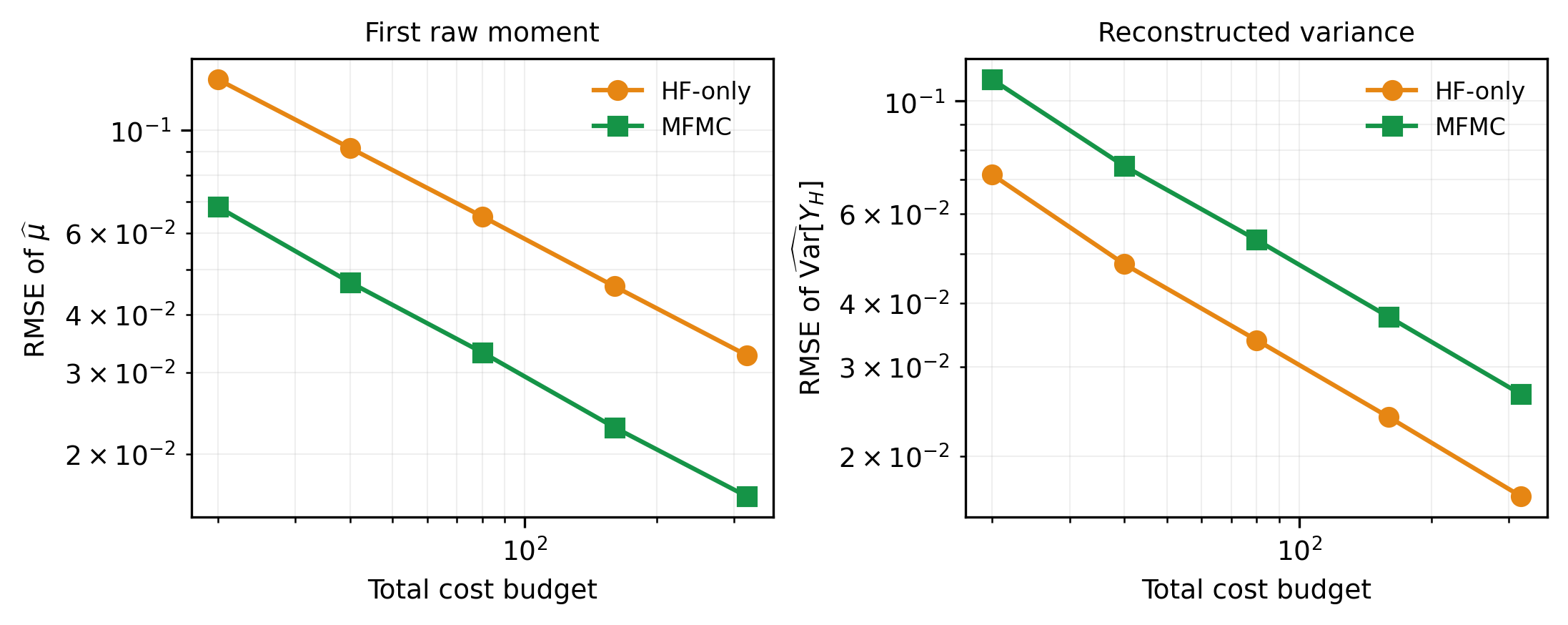}
  \caption{Equal-cost performance in the analytic example. Sample counts chosen from the unsquared-response correlation improve estimation of the first raw moment (left). Because the squared responses are only weakly correlated, the reconstructed variance performs worse than HF-only Monte Carlo (right). The two raw moments use separate weights but the same nested sample counts.}
  \label{fig:toy_moment_performance}
\end{figure}

To show how this affects the estimator, we set the relative costs to $c_H=1$ and $c_L=\ToyLFRelativeCost$. We choose the sample counts from the correlation between the unsquared responses, as in the main study. For this example with one LF model, the continuous ratio is $N_L/N_H=\ToyAllocationRatio$. The mean and second raw moment use separate optimal control-variate weights but share these sample counts. Figure~\ref{fig:toy_moment_performance} shows the RMSE from \ToyPerformanceRepeats{} independent repetitions at the same total cost. At the largest budget, $B=\ToyLargestBudget$, MFMC uses $N_H=\ToyLargestMFMCNHigh$ and $N_L=\ToyLargestMFMCNLow$. It reduces the mean RMSE by a factor of \ToyLargestMeanGain, but its reconstructed-variance RMSE is \ToyLargestVariancePenalty{} times the HF-only RMSE. The strong response correlation gives efficient sample counts for $\E[Y_H]$. However, those counts include too few HF samples to compensate for the weak correlation between squared responses. This example shows why $\rho_f$, $\rho_g$, and the reconstructed variance must be assessed separately.

\FloatBarrier
\subsection{Benchmark Setup and Evaluation}
The benchmark uses the Cube, scaled GOCE-like, and scaled CHAMP-like surfaces after they passed the mesh, domain, time-step, particle-number, sampling-window, force-sign, and reference-area checks. Only then were the uncertain-input outputs included in the analysis.

Table~\ref{tab:benchmark-cases} summarizes the purpose of each case. The four Cube cases test the same canonical geometry and 12-dimensional input definition at different altitudes. The GOCE-like and CHAMP-like cases test both MFMC estimators on more complex shapes and on environmental states whose variables retain their empirical dependence. These cases test changes in geometry and input distribution; they are not independent flight-validation cases.

\begin{table*}[!htbp]
\centering
\small
\caption{Role and scope of the six retained benchmark cases.}
\label{tab:benchmark-cases}
\begin{tabular}{llcrrl}
\toprule
Case & benchmark role & dimension & pilot & estimator budgets [CPU h] & HF reference \\
\midrule
Cube, 100 km & canonical altitude case & 12 & 32 & 195--973 (3) & 450 \\
Cube, 200 km & canonical altitude case & 12 & 32 & 91--457 (3) & 450 \\
Cube, 300 km & canonical altitude case & 12 & 32 & 241--1204 (3) & 450 \\
Cube, 400 km & canonical altitude case & 12 & 32 & 212--1062 (3) & 450 \\
GOCE-like, 244 km & trajectory/geometry transfer & row + 9 & 50 & 16--163 (4) & 450 \\
CHAMP-like, 358 km & trajectory/geometry transfer & row + 9 & 50 & 24--238 (4) & 450 \\
\bottomrule
\end{tabular}
\end{table*}

Each uncertain-input case begins with a paired DSMC--TPMC--Sentman pilot. The Cube pilots contain 32 paired samples and the trajectory-conditioned GOCE-like and CHAMP-like pilots contain 50 paired samples. The pilot estimates response and squared-response correlations, moment-specific weights, and measured cost ratios. Ten independent estimator repetitions are evaluated at each complete computational budget.

For TPMC and Sentman, the pilot record reports the correlations for $C_D$ and $C_D^2$, the fitted weights, sample-count ratios, and the frequencies of negative or unstable weights. The sample counts do not depend on the centered-square correlations. We calculated these correlations after the main analysis to investigate the GOCE-like variance failure.

Table~\ref{tab:benchmark-costs} reports the median measured cost per evaluation. One TPMC evaluation costs 0.115--0.395 times as much as one DSMC evaluation, whereas one Sentman evaluation costs $4.82\times10^{-6}$--$1.36\times10^{-4}$ times as much. The table also reports the total measured cost of all stored pilot evaluations, expressed in equivalent DSMC runs. This total ranges from 35.15 to 59.45 DSMC-equivalent evaluations. The primary comparisons include only estimation cost because the same pilot supports every tested budget. A separate sensitivity analysis adds the full pilot cost to each comparison. The analysis outputs provide the sample counts and their ranges across repetitions. At the smallest budgets, only one or two HF evaluations are available.

The allocation routine first enforces the configured minimum number of DSMC runs on the continuous sample counts. It then rounds the counts and adjusts the simulation hierarchy to satisfy the measured-cost budget. This final adjustment can reduce the realized DSMC count below the nominal minimum. As a result, some stored trajectory estimates contain only one DSMC evaluation. We include these estimates because excluding them after seeing the results would change the prespecified benchmark. We report their instability and repeat the comparison using only estimates with at least five DSMC evaluations. Seven case--budget combinations meet this condition. All seven have a mean-drag gain above one, while four have a variance gain above one. This sensitivity analysis supports the conclusion for the mean but shows that the smallest DSMC sample counts provide weak evidence about variance performance.

\begin{table*}[!htbp]
\centering
\caption{Measured median per-evaluation costs and reconstructed realized pilot costs. Cost ratios are relative to one DSMC evaluation. For each pilot sample, the reported pilot cost adds the recorded costs of its DSMC, TPMC, and Sentman evaluations and then sums over all pilot samples.}
\label{tab:benchmark-costs}
\resizebox{\textwidth}{!}{%
\begin{tabular}{lrrrrrr}
\toprule
Case & $N_{\rm pilot}$ & DSMC [CPU h] & TPMC/DSMC & Sentman/DSMC & pilot [CPU h] & pilot [HF equiv.] \\
\midrule
Cube, 100 km & 32 & 17.6 & 0.324 & 7.91e-06 & 807 & 45.74 \\
Cube, 200 km & 32 & 9.02 & 0.331 & 1.29e-05 & 395 & 43.85 \\
Cube, 300 km & 32 & 27.5 & 0.224 & 3.92e-06 & 968 & 35.15 \\
Cube, 400 km & 32 & 20.8 & 0.395 & 4.82e-06 & 948 & 45.51 \\
GOCE-like, 244 km & 50 & 7.82 & 0.14 & 0.000136 & 465 & 59.45 \\
CHAMP-like, 358 km & 50 & 11.7 & 0.115 & 1.28e-05 & 656 & 56.26 \\
\bottomrule
\end{tabular}}
\end{table*}

For every accepted case, pilot and estimator samples are disjoint. The primary direct targets are
\begin{equation}
  \mu=\E_{\Pi_k}[\CD],\qquad m_2=\E_{\Pi_k}[\CD^2],
\end{equation}
and the derived standard deviation is
\begin{equation}
  \sigma_{\CD}=\sqrt{m_2-\mu^2}.
\end{equation}
The reference mean and population variance use 450 physically distinct HF responses per case. These sets combine pilot and estimator evaluations with additional HF runs and are therefore not independent of every estimator repetition. The reported RMSE measures convergence within this dataset, not error against an external validation dataset. Negative finite-sample variance estimates are included without clipping.

At budget $B$, the equal-cost DSMC-only comparator receives $N_H=\lfloor B/\bar c_H\rfloor$ samples, where $\bar c_H$ is the measured pilot mean cost. Its expected mean RMSE is $s_H/\sqrt{N_H}$. For the biased raw-moment variance estimator $N_H^{-1}\sum_i C_{D,i}^2-(N_H^{-1}\sum_i C_{D,i})^2$, the expected MSE is calculated from the second and fourth central moments of the 450-run HF reference distribution, including its finite-$N_H$ bias. Thus the HF-only curves are expected Monte Carlo errors, whereas the multifidelity curves are empirical RMSE values over ten repetitions.

We constructed the TPMC-only estimator after inspecting the results; it was not a prospectively optimized design. For every case and budget, it uses all stored HF and TPMC evaluations and the original TPMC weights for each moment. It removes only the Sentman corrections and assigns the same estimation budget as DSMC--TPMC--Sentman MFMC. Because Sentman is extremely cheap, this slightly overstates the relative cost of the TPMC-only estimator. The comparison shows the statistical effect of adding Sentman without generating new samples or changing the HF sample count.

\paragraph{Reproducibility record.}
For each campaign, we store the complete configuration and its fingerprint, the global seed, the seed derived for each cell, the sampled inputs, model outputs, weights for each moment, sample counts, and measured costs. Seeds are generated deterministically from the global seed and the identifiers for the study, case, budget, repetition, model, and evaluation. Each production repetition therefore has a distinct recorded seed. The Cube DSMC and TPMC jobs use 128 MPI ranks. The trajectory DSMC jobs use 128 ranks for GOCE-like and 256 ranks for CHAMP-like. The analysis scripts record the bootstrap seed and number of repetitions. The environment manifest records Python 3.12.3 and the exact versions of the post-processing packages. Historical records identify the solver adapters but not the exact versions of PICLas, ADBSat, the compiler, or MPI. These versions remain unresolved, and we do not claim bitwise reproduction of the solver results.

\section{Results}
\label{sec:results}
Table~\ref{tab:hf-reference-targets} summarizes the finite HF reference sets. Each case contains 450 physically distinct \textsc{PICLas} DSMC drag coefficients. For the Cube cases, the reference mean increases from 2.3946 at 100 km to 2.5825 at 400 km, while the standard deviation remains between 0.0534 and 0.0681. The reference means are 4.0016 for the GOCE-like case and 3.3456 for the CHAMP-like case; their standard deviations are 0.1912 and 0.2406. Across all cases, the standard error of the reference mean ranges from 0.00252 to 0.01136. RMSE differences of this size should not be interpreted as differences relative to an exact population moment.

\begin{table}
\centering
\caption{HF reference statistics from 450 physically distinct PICLas DSMC evaluations per case.}
\label{tab:hf-reference-targets}
\begin{tabular}{lrrrrr}
\toprule
Case & $N_{HF}$ & $E[C_D]$ & $\mathrm{Var}(C_D)$ & $\sigma(C_D)$ & SE$(E[C_D])$ \\
\midrule
Cube, 100 km & 450 & 2.3946 & 0.0028481 & 0.053368 & 0.0025186 \\
Cube, 200 km & 450 & 2.5136 & 0.0046313 & 0.068054 & 0.0032117 \\
Cube, 300 km & 450 & 2.5498 & 0.004443 & 0.066656 & 0.0031457 \\
Cube, 400 km & 450 & 2.5825 & 0.0036001 & 0.060001 & 0.0028316 \\
GOCE-like, 244 km & 450 & 4.0016 & 0.036552 & 0.19118 & 0.0090226 \\
CHAMP-like, 358 km & 450 & 3.3456 & 0.057893 & 0.24061 & 0.011355 \\
\bottomrule
\end{tabular}
\end{table}

\subsection{RMSE of the Mean Drag Coefficient}
\begin{figure}[!htbp]
  \centering
  \includegraphics[width=\textwidth]{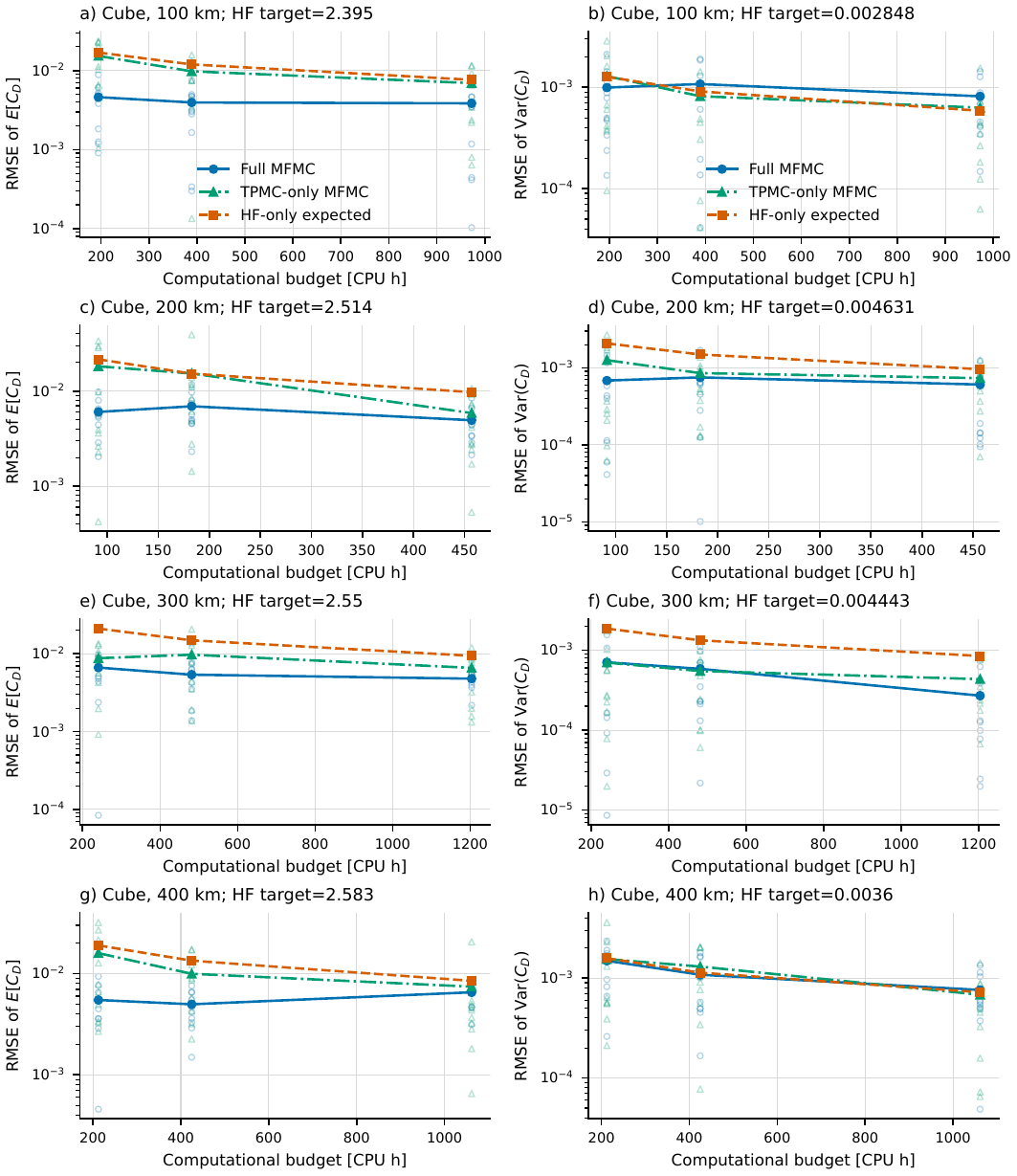}
  \caption{RMSE against the 450-run HF reference for the four Cube cases. The left column shows $\E[\CD]$ and the right column shows the reconstructed $\Var(\CD)$. Lines show RMSE over ten repetitions for DSMC--TPMC--Sentman MFMC and TPMC-only MFMC, together with the expected error when the complete estimation budget is spent on DSMC; faint symbols show absolute errors of individual multifidelity repetitions.}
  \label{fig:rmse_cube_cases}
\end{figure}

\begin{figure}[!htbp]
  \centering
  \includegraphics[width=\textwidth]{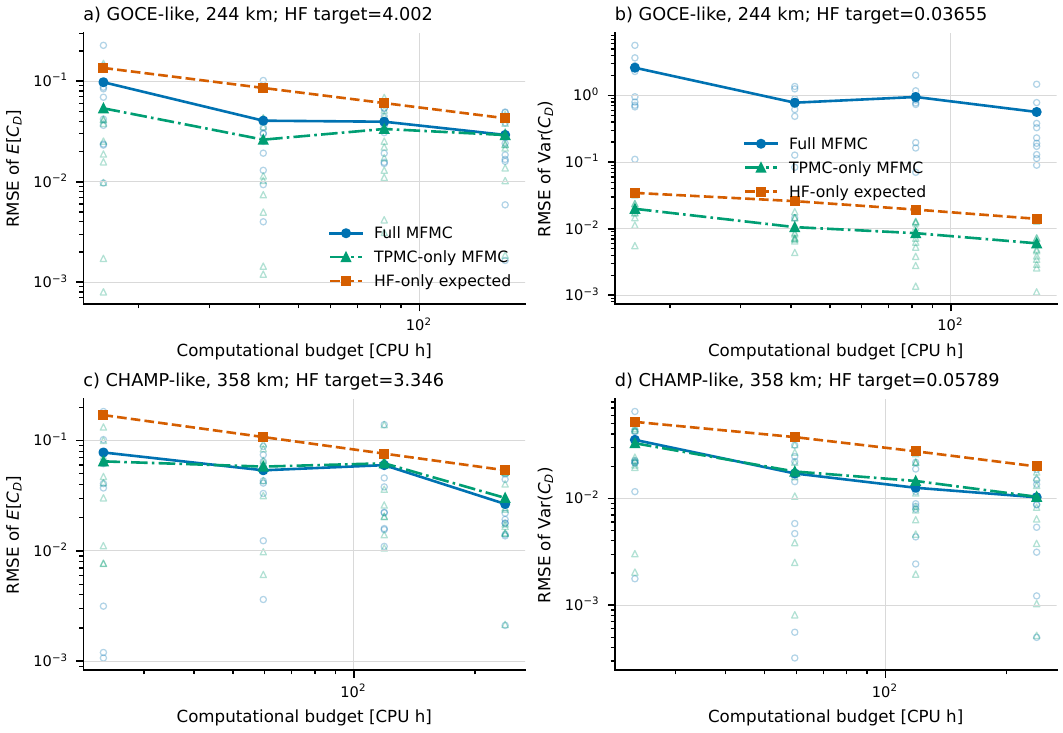}
  \caption{RMSE against the 450-run HF reference for the trajectory-conditioned GOCE-like and CHAMP-like cases. Lines show RMSE over ten repetitions for DSMC--TPMC--Sentman MFMC and TPMC-only MFMC, together with the expected error when the complete estimation budget is spent on DSMC; faint symbols show individual multifidelity errors. Negative variance estimates contribute through their signed error before squaring and are not clipped.}
  \label{fig:rmse_trajectory_cases}
\end{figure}

Figures~\ref{fig:rmse_cube_cases} and~\ref{fig:rmse_trajectory_cases} compare DSMC--TPMC--Sentman MFMC, TPMC-only MFMC, and equal-budget HF-only estimation against the corresponding 450-run HF references. For an estimator $e$, define the improvement factor
\begin{equation}
  G_e=\frac{\mathrm{RMSE}_{\mathrm{HF-only}}}{\mathrm{RMSE}_{e}}.
\end{equation}
DSMC--TPMC--Sentman MFMC has a lower mean-drag RMSE than HF-only Monte Carlo in every case and budget. Its median $G_e$ for each case ranges from 1.50 for GOCE-like to 3.03 for Cube at 100 km (Table~\ref{tab:rmse-performance-ratio}). TPMC-only MFMC improves 19 of the 20 comparisons and has median gains of 1.10--2.16. The three-model estimator has a larger median gain than TPMC-only in five cases. For GOCE-like, TPMC-only is better, with a gain of 2.16 instead of 1.50. Sentman therefore usually improves the mean estimate, but not for every geometry.

We used a descriptive bootstrap to assess variation across the ten full-hierarchy repetitions. For each case and budget, we resampled the ten errors 100,000 times and kept the analytic HF-only RMSE fixed. The 5th percentile of the gain is above one in 18 of the 20 mean-drag comparisons. The exceptions are the smallest GOCE-like budget and the third CHAMP-like budget. These intervals describe uncertainty from the finite number of repetitions for a fixed pooled HF reference and estimated HF error distribution. They do not include the uncertainty in that reference or its dependence on the estimator samples.

\subsection{Reconstructed Drag-Coefficient Variance}
At every tested budget, DSMC--TPMC--Sentman MFMC reduces the RMSE of the reconstructed variance for Cube at 200 and 300 km and for CHAMP-like. The median gains for these cases are 1.98, 2.64, and 2.07. Results are mixed for Cube at 100 km and approximately equal to HF-only Monte Carlo for Cube at 400 km. TPMC-only gives the same overall pattern. Its median gains are 1.65, 2.42, and 1.91 for Cube at 200 and 300 km and for CHAMP-like, and it remains close to HF-only Monte Carlo for Cube at 100 and 400 km. Sentman therefore provides a modest additional benefit in the three consistently successful cases, but does not produce clear gains in the other two Cube cases.

For the variance, the 5th percentile of the bootstrapped gain is above one in 10 of the 20 case--budget combinations. It is below one for all four GOCE-like combinations and for two Cube combinations at 100 km. All intervals for Cube at 400 km include one. The gains for Cube at 200 and 300 km and for CHAMP-like are therefore stable across the ten repetitions, whereas comparisons close to one remain inconclusive.

The clearest failure occurs for the GOCE-like variance. Its median gain is 0.023, with values from 0.013 to 0.033, so HF-only Monte Carlo has a lower variance RMSE at every tested budget. Nineteen of the 40 GOCE-like MFMC variance estimates are negative. At the largest budget, the MFMC variance RMSE is 0.566, compared with an expected value of 0.0141 for HF-only Monte Carlo at the same cost. This result is consistent with cancellation error in $\widehat{m}_2-\widehat{\mu}^{2}$. Because the two raw moments are corrected separately, a small mismatch between them can dominate the reference variance of 0.0366. The GOCE-like mean estimate remains useful: its median MFMC gain is 1.50. However, the sample counts chosen for the mean and the subsequent raw-moment subtraction do not give a reliable GOCE-like variance estimate.

\subsection{Exploratory GOCE-like Failure Diagnosis}
\begin{figure}[!htbp]
  \centering
  \includegraphics[width=\textwidth]{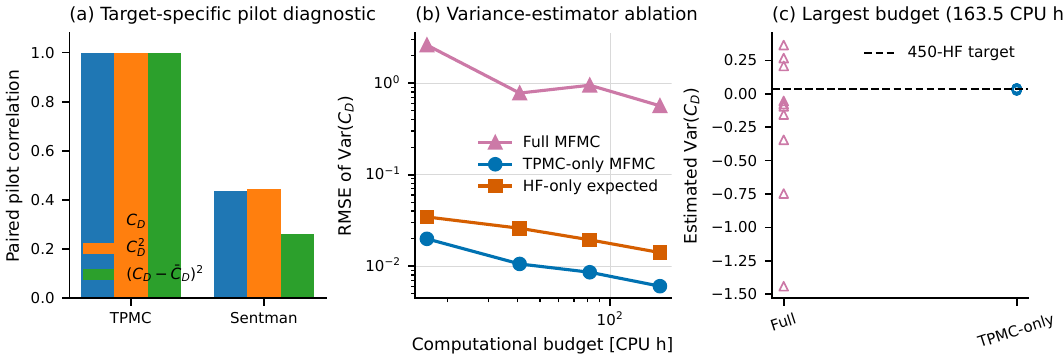}
  \caption{Exploratory analysis of the GOCE-like variance failure. (a) Correlations between paired DSMC and low-fidelity outputs in the 50-sample pilot for the response, its square, and its squared deviation from the mean. (b) RMSE of the reconstructed variance for DSMC--TPMC--Sentman MFMC, the estimator without Sentman, and HF-only Monte Carlo at the same budget. (c) Individual variance estimates at the largest budget; the dashed line is the 450-HF reference. The estimator without Sentman uses all stored HF and TPMC evaluations and removes only the Sentman corrections.}
  \label{fig:goce-variance-diagnostic}
\end{figure}

The 50 paired GOCE-like pilot samples show a clear difference between TPMC and Sentman (Fig.~\ref{fig:goce-variance-diagnostic}a). The TPMC correlations with DSMC are 0.9988 for $\CD$, 0.9989 for $\CD^2$, and 0.9994 for the squared deviations from the mean. The corresponding Sentman correlations are only 0.4349, 0.4445, and 0.2612. Despite these lower correlations, the sample counts chosen for mean estimation include 221--1458 Sentman evaluations but only one or two HF evaluations. DSMC--TPMC--Sentman MFMC still improves the GOCE-like mean over HF-only Monte Carlo, although its median gain is smaller than that of TPMC-only. For the variance, the separately weighted Sentman corrections do not preserve the relation between the raw-moment errors that is needed for their subtraction.

We tested this explanation by removing Sentman after the main analysis. The resulting estimator uses every stored HF and TPMC evaluation and the original TPMC weight for each moment, but omits the Sentman correction from both raw moments. We did not remove or regenerate any samples. Across the four budgets, the TPMC-only variance RMSE decreases from 0.0199 to 0.00603, compared with 2.61 to 0.566 for DSMC--TPMC--Sentman MFMC (Fig.~\ref{fig:goce-variance-diagnostic}b). Relative to HF-only Monte Carlo at the same cost, the gain ranges from 1.73 to 2.46, with a median of 2.30. None of the 40 TPMC-only variance estimates is negative (Fig.~\ref{fig:goce-variance-diagnostic}c). These results support the explanation that Sentman's weak correlation for the variance-related quantities caused the failure. Because we removed Sentman after examining the GOCE-like results, this is evidence for how a future estimator might be designed, not a confirmatory performance result.

\begin{table*}[!htbp]
\centering
\small
\caption{Ratio $\mathrm{RMSE}_{\mathrm{HF-only}}/\mathrm{RMSE}_{\mathrm{estimator}}$ across computational budgets. Values are median [minimum, maximum].}
\label{tab:rmse-performance-ratio}
\resizebox{\textwidth}{!}{%
\begin{tabular}{lllll}
\toprule
Case & Full: $E[C_D]$ & Full: $\mathrm{Var}(C_D)$ & TPMC-only: $E[C_D]$ & TPMC-only: $\mathrm{Var}(C_D)$ \\
\midrule
Cube, 100 km & 3.03 [1.98, 3.67] & 0.85 [0.72, 1.28] & 1.10 [1.10, 1.22] & 0.98 [0.93, 1.11] \\
Cube, 200 km & 2.20 [1.97, 3.57] & 1.98 [1.58, 3.03] & 1.18 [1.00, 1.66] & 1.65 [1.31, 1.75] \\
Cube, 300 km & 2.77 [1.97, 3.17] & 2.64 [2.28, 3.13] & 1.53 [1.43, 2.40] & 2.42 [1.95, 2.66] \\
Cube, 400 km & 2.70 [1.30, 3.46] & 1.05 [0.95, 1.06] & 1.19 [1.14, 1.35] & 1.02 [0.87, 1.06] \\
GOCE-like, 244 km & 1.50 [1.39, 2.12] & 0.02 [0.01, 0.03] & 2.16 [1.47, 3.27] & 2.30 [1.73, 2.46] \\
CHAMP-like, 358 km & 2.01 [1.27, 2.18] & 2.07 [1.47, 2.20] & 1.82 [1.22, 2.63] & 1.91 [1.58, 2.10] \\
\bottomrule
\end{tabular}}
\end{table*}

Three patterns summarize the results. First, DSMC--TPMC--Sentman MFMC improves the mean estimate in every case and budget. Second, variance gains depend on the case and are less consistent. Third, the GOCE-like variance failure occurs when the Sentman correction is included and disappears when that correction is removed.

\FloatBarrier

\section{Discussion}
\label{sec:discussion}

\subsection{MFMC Efficiency for the Mean Drag Coefficient}
The results give a clear answer for the mean. At the same estimation cost, DSMC--TPMC--Sentman MFMC improves $\E[\CD]$ throughout the benchmark, and TPMC-only improves all but one comparison. A biased collisionless model can therefore be an effective control variate if its changes across the sampled inputs follow the DSMC response. Adding Sentman helps only when it is sufficiently correlated with DSMC for the case and quantity being estimated.

The gain $G_e$ compares errors at the same estimation cost; it is not a direct reduction in wall-clock time. Under the ideal Monte Carlo relation $\mathrm{RMSE}\propto N^{-1/2}$, reproducing an error reduction of $G_e$ with HF-only Monte Carlo would require approximately $G_e^2$ times as many independent HF samples. This conversion is only a scaling reference. If the full pilot cost is assigned to one estimate, a total-cost gain remains only for the Cube cases. Reusing the pilot across several statistics or budgets is therefore important for practical efficiency.

\subsection{MFMC Performance for Variance Estimation}
The answer for the reconstructed variance is conditional. Both MFMC estimators improve three cases and perform close to HF-only Monte Carlo in two cases. For GOCE-like, DSMC--TPMC--Sentman MFMC fails, whereas TPMC-only improves the estimate. This difference agrees with the analytic example and the delta-method calculation: variance performance depends on the correlations between squared responses and on the covariance between the two raw-moment errors, not only on $\rho_{f,\ell}$. Whether a low-fidelity model is useful therefore depends on the model, input distribution, and quantity being estimated.

The GOCE-like results illustrate this dependence. TPMC is strongly correlated with DSMC for the response, its square, and the squared deviation from the mean; Sentman is not. At the same time, the sample counts chosen for the mean include only one or two HF evaluations and many inexpensive Sentman evaluations. The failure is therefore consistent with three combined effects: weak Sentman correlations for the variance-related quantities, too few HF samples, and error cancellation in $\widehat{m}_2-\widehat{\mu}^{2}$. The exploratory removal of Sentman supports this explanation, but it does not establish a universal correlation threshold or provide a confirmatory performance result.

Before reconstructing a variance, each low-fidelity model should be checked for that specific task. The checks should include its correlations with the response, squared response, and squared deviation from the mean. The design should also require a minimum number of HF samples and examine the two raw-moment errors together. The GOCE-like result remains exploratory until a rule for including or excluding a model is defined in advance and tested on independent data or by cross-fitting.

\subsection{Interpretation of the Results}
The result answers the initial question: at a fixed estimation budget, both MFMC estimators obtain more accurate mean estimates than DSMC alone. This can allow an expensive rarefied-flow study to cover more uncertain inputs and provide both a drag estimate and its uncertainty for later orbit propagation. The failure for the variance gives a separate lesson: good performance for a linear quantity does not justify using the same models and sample counts for a nonlinear quantity. We therefore report both the failure of DSMC--TPMC--Sentman MFMC and the improvement obtained after removing Sentman. We neither clip the negative estimates nor present the later model removal as prespecified. A validation based only on the mean would have missed this problem.

The reported $\Var(\CD)$ combines variation in the prescribed physical inputs with variation between DSMC particle seeds. It is therefore not a separate estimate of physical drag uncertainty or a mission-level risk distribution. Section~\ref{sec:scope} gives the resulting limits on physical and aerodynamic interpretation.

\section{Scope and Limitations}
\label{sec:scope}
The benchmark estimates the DSMC distribution produced by the prescribed physical-input distributions and particle-sampling procedure. It is not a flight validation, material calibration, or orbit-propagation study, but it provides a methodological basis for using multifidelity aerodynamic uncertainty estimates in such applications. The scaled GOCE-like and CHAMP-like surfaces demonstrate the application of the method to geometries beyond the canonical Cube case. The independently sampled input perturbations are not a posterior distribution fitted to thermospheric data. The single shared value of $\alpha_E$ also does not describe material ageing, contamination, exposure to atomic oxygen, or separate normal and tangential accommodation.

The HF reference values are Monte Carlo estimates based on 450 runs, not exact population moments. They combine physically distinct pilot and estimator evaluations with additional HF runs and are therefore not fully independent of the estimators used to calculate the RMSE. The reported gains show convergence within this dataset at equal budgets. The negative GOCE-like variance estimates also demonstrate that an accurate mean estimate does not guarantee an accurate nonlinear variance estimate. The exploratory TPMC-only result must be confirmed with a model-selection threshold defined in advance or in an independent study. A stronger validation would use an independent HF reference set and compare with a direct non-negative variance estimator or an estimator that controls the two raw moments jointly.

The empirical RMSE at each budget is based on ten estimator repetitions. The gains therefore describe this benchmark and do not provide precise uncertainty intervals for a broader population. The primary comparisons also include only estimation cost. The paired pilot is an initial cost, and its practical effect depends on how many later analyses can reuse it. These limitations matter especially when two gains are close or when the observed ratios are applied to a new geometry or input distribution.

\section{Conclusion}
\label{sec:conclusion}
Drag predictions for satellites in very low Earth orbit depend on an atmosphere that is rarefied, multi-species and strongly variable, so a single, deterministic drag coefficient is not meaningful: orbit maintenance, lifetime prediction and propellant budgeting need the mean and the spread of $C_\mathrm{D}$ under that variability. DSMC can provide these statistics with the right physics, but a Monte Carlo study built on DSMC alone is expensive, because every input sample requires a new particle simulation. This paper asked whether the cost can be reduced by pairing DSMC with two much cheaper collisionless models, TPMC and the Sentman panel model, as control variates in a multifidelity Monte Carlo estimator. Two questions were posed: whether such a hierarchy lowers the error of DSMC-based drag statistics at equal cost across different geometries, and whether a gain for the mean carries over to a variance reconstructed from two separately estimated moments.

The answer to the first question is yes for the mean. In all six benchmark cases and at every budget tested, the DSMC--TPMC--Sentman estimator reduced the RMSE of the mean drag coefficient relative to DSMC-only Monte Carlo, by median factors of 1.50 to 3.03. Because Monte Carlo error falls with the square root of the sample count, these factors correspond to roughly two to nine times fewer DSMC runs for the same accuracy. The gain is largest where the cheap models track DSMC most closely, and it is also present at 100 km, where collisions matter and the collisionless models are physically least justified. A biased model can thus be a useful control variate as long as its fluctuations across the sampled inputs follow those of DSMC. The TPMC-only estimator improved 19 of 20 comparisons, and the Sentman branch added further benefit in five of the six cases. The benefit depends on reusing the pilot: if its full cost is charged to a single estimate, only the Cube cases remain ahead of DSMC-only Monte Carlo.

The answer to the second question is no, not automatically. For three cases the hierarchy improved the variance by factors of about two, for two Cube cases it was close to parity, and for the GOCE-like case it failed outright: the variance error exceeded that of DSMC-only Monte Carlo at every budget, and nearly half the estimates were negative. The pilot diagnostics explain why. TPMC is correlated with DSMC above 0.998 for the response, its square and its squared deviation from the mean, whereas the Sentman model reaches only 0.26 to 0.44 for the GOCE-like geometry. A mean-oriented allocation then leaves one or two DSMC runs per estimate and spends the budget on a model that is uninformative for the variance. Removing the Sentman correction, with all other samples and weights unchanged, restored a variance gain of 1.7 to 2.5 at every budget. The analytic example of Section 5 shows that the same decoupling between the response and its square can occur in a simple model, so the GOCE-like result is a general caution rather than a peculiarity of that geometry.

For practitioners, the conclusions are therefore the following. Multifidelity control variates are a reliable way to obtain the mean DSMC drag coefficient under atmospheric and attitude variability at a fraction of the DSMC-only cost, provided one pilot serves several budgets or statistics. For the variance, each low-fidelity model must be screened separately for the statistic it is meant to improve, using the correlations of the response, its square and its centred square, and the allocation should guarantee a minimum number of DSMC runs. A model that is excellent for the mean can be detrimental to the variance.

The study has clear limits. The references are internal 450-run DSMC estimates, the geometries are scaled benchmarks, the input distributions are broad test ranges rather than calibrated models, and the decision to remove the Sentman model was taken after inspecting the results. Future work should fix the screening rule in advance and test it against an independent DSMC reference, compare the two-moment reconstruction with a direct variance estimator at the same cost, and apply the hierarchy at flight scale so that the resulting drag distribution can be carried through to orbit lifetime and propellant margins.

\section*{CRediT authorship contribution statement}
Jovan Boskovic: Conceptualization, Methodology, Software, Validation, Formal analysis, Investigation, Data curation, Visualization, Writing -- original draft, Writing -- review \& editing. Marcel Pfeiffer: Conceptualization, Methodology, Supervision, Writing -- review \& editing. Andrea Beck: Conceptualization, Methodology, Supervision, Funding acquisition, Project administration, Writing -- review \& editing.

\section*{Declaration of competing interest}
The authors declare that they have no known competing financial interests or personal relationships that could have appeared to influence the work reported in this paper.

\section*{Declaration of generative AI and AI-assisted technologies in the\\writing process}
During the preparation of this work, the authors used OpenAI Codex to assist with consistency checks, LaTeX preparation, and reproducible post-processing code. Scientific-writing and evidence-reconciliation procedures followed \citet{Kassis2026}. The authors reviewed and edited all affected content and take full responsibility for the publication.

\section*{Data and code availability}
The reproducibility package contains the six configuration snapshots, sampled inputs, model outputs, sample counts and weights, the 2,700-row table identifying the HF reference samples, derived analysis tables, the environment manifest, and the scripts needed to reproduce the reported tables and figures. A permanent archival DOI will be provided before submission. The stored campaign configurations do not contain the exact historical versions of the solvers or MPI; the manifest marks these versions as unresolved.

\section*{Funding}
This work was funded by the Deutsche Forschungsgemeinschaft (DFG, German Research Foundation) -- Project-ID 516238647 -- SFB 1667/1 (ATLAS - Advancing Technologies of Very Low-Altitude Satellites). The funder had no role in study design; data collection, analysis, or interpretation; manuscript preparation; or the decision to submit the article for publication.

\section*{Acknowledgements}
The authors thank the Leibniz Supercomputing Centre for granting compute time on SuperMUC--NG under project no. 11245. We acknowledge EuroHPC Joint Undertaking for awarding us access to HPC Vega, Slovenia.

\bibliographystyle{cas-model2-names}
\bibliography{cas-refs}

\end{document}